\documentstyle[12pt]{article}
\def\ka{\kappa}

\def\f{\varphi}

\def\w{\omega}
\def\la{\lambda}
\def\ds{\zeta}
\def\mo{m_{1}}
\def\mt{m_{2}}
\def\d{\partial}
\def\fo{\f_{1}}
\def\zo{z_{1}}
\def\zt{z_{2}}
\def\nt{n_{2}}
\def\no{n_{1}}

\def\e{\mbox{e}}
\def\ft{\f_{2}}
\def\fn{\f_{i}}
\def\fj{\f_{j}}
\def\eps{\epsilon}
\def\ds{\zeta}
\def\dso{\zeta_{1}}
\def\dst{\zeta_{2}}
\begin{document}
\begin{titlepage}
\title{Zeros of tree amplitudes at rest and symmetries of mechanical systems}
\author{
  M.V. Libanov, V.A. Rubakov and S.V. Troitsky\\
{\small{\em Institute for Nuclear Research of the Russian Academy of Sciences,
}}\\
{\small{\em 60th October Anniversary prospect, 7a, Moscow 117312}}\\
}
\end{titlepage}
\maketitle

\begin{center}
{\bf Abstract}
\end{center}
We consider the tree amplitudes of production of $n_2$ scalar particles by
$n_1$ particles of another kind, where both initial and final particles are
at rest and on mass shell, in a model of two scalar fields with $O(2)$
symmetric interaction and unequal masses. We find that these amplitudes
are zero except for the lowest possible $n_1$ and $n_2$, and that the
cancellation of the corresponding Feynman graphs occurs due to a special
symmetry of the classical mechanical counterpart of this theory.
This feature is rather general and is inherent in various other scalar
field theories.

\newpage
{\bf 1}.
Recently, powerful techniques have been developed for
calculating tree amplitudes of the production of $n$
scalar particles at $n$-particle treshold by one or two
initial particles (virtual or real)
\cite{Voloshin,Brown,Kleiss,Voloshin',Voloshin0,Kleiss0,BZh,we}.
The implementation of these techniques revealed an interesting property that
some of the amplitudes are equal to zero. One type of the
cancellation between the tree graphs occurs for
processes involving two on-shell initial particles with
non-zero spatial momenta and sufficiently large number of
final particles, all of which are at rest.  For example, in
the theory of one scalar field with $\f^{4}$
self-interaction, the tree amplitudes of the scattering of two particles
into $n$ particles at the threshold
vanish for $n>2$ in the case of unbroken reflection symmetry and for $n>4$
in the case of broken symmetry \cite{Voloshin',Voloshin0}. This property
is related to the reflectionlessness of certain
potentials in quantum mechanics \cite{Voloshin0,Kleiss0}.
Different type of the zeros of the tree amplitudes at rest was found
 in ref.\cite{we}.
The processes of the production of $n_{1}$ particles of the field $\fo$
and $n_{2}$ particles of the field $\ft$ by one initial particle
were considered in the framework of the model with
$O(2)$-symmetric interaction and unequal masses, whose
lagrangian is

\begin{equation}
L=\frac{1}{2}(\d_{\mu}
\fo)^{2}+\frac{1}{2}(\d_{\mu}\ft)^{2}-\frac{m_{1}^{2}}{2}\fo^{2}-
\frac{m_{2}^{2}}{2}\ft^{2}-
\la(\fo^{2}+\ft^{2})^{2}
\label{1*}
\end{equation}
In the case of broken reflection symmetry
       $\fo \to -\fo$,
the production of $n$ particles $\ft$ by one particle $\fo$,
where both initial and final particles are at rest and
on-shell, is kinematically allowed under certain conditions on $\mo$ and $\mt$.
It has been observed that the tree amplitude of this process
vanishes at $n>2$. No explanation of this property was
given in ref.\cite{we}, although it was suspected that it might
be related to a specific symmetry of the interaction
$(\fo^{2}+\ft^{2})^{2}$.

 In this paper we extend this result of ref.\cite{we} and show that the tree
amplitudes of
the processes of the production of $\nt$ particles $\ft$ by
$\no$ particles $\fo$, all at rest and on-mass-shell, where
$n_{1}$ and $n_{2}$ are coprime numbers up to one common
divisor 2, vanish in the model (\ref{1*}) except for the
cases $n_{1}=n_{2}=2$ when the reflection symmetry is unbroken and $n_{1}=1$,
 $n_{2}=2$ when the symmetry is broken.
We also relate the cancellation of the corresponding tree diagrams to the
integrablility of the classical mechanical system with the
hamiltonian
\begin{equation}
H=\frac{1}{2}(\dot{\fo})^{2}+\frac{1}{2}(\dot{\ft})^{2}+
\frac{m_{1}^{2}}{2}\fo^{2}+\frac{m_{2}^{2}}{2}\ft^{2}+
\la(\fo^{2}+\ft^{2})^{2}
\label{1a*}
\end{equation}
which is obtained from eq.(\ref{1*}) by discarding the
space dependence of $\fo$ and $\ft$.  Namely, we show
explicitly how the non-trivial symmetry \cite{Ind} of the
system (\ref{1a*}) (which is the simplest case of
the Garnier systems \cite{Perelomov}) leads to the
nullification of the tree amplitudes at rest.

{\bf 2}. For calculating the tree amplitudes at the threshold, two methods have
been
employed. One of them is based on recursion relations between diagrams with
different numbers of final particles \cite{Voloshin}, and the other makes use
of classical field equations with special boundary conditions \cite{Brown}.
Let us extend the classical solution method to the case when the initial
particles are also on mass shell and at rest.

Let us consider the model of two scalar fields, $\fo$ and $\ft$, with quartic
interaction term,
\[
V(\fo,\ft)=\frac{m_{1}^{2}}{2}\fo^{2}+
\frac{m_{2}^{2}}{2}\ft^{2}+\sum_{i,k=1,2}\la_{ik}
\f_{i}^{2}\f_{k}^{2}\]
Our purpose is to calculate the tree amplitudes of the
production of $n_{2}$ particles $\ft$ by $\no$ particles
$\fo$, where all initial and final particles are on-shell
and their spatial momenta are zero. This process is allowed
 by energy conservation when $n_{1}m_{1}=n_{2}m_{2}$. We
 study the case when no disconnected diagrams exist, so we
keep $n_{1}$ and $n_{2}$ coprime up to one common divisor
2.

\newpage
The LSZ reduction formula for connected amplitudes reads,
\begin{equation}
\begin{array}{l}
\displaystyle <n_{2},\ft |n_{1}, \fo>= \\
\displaystyle
\prod\limits_{a=1}^{n_{2}}\,
\prod\limits_{b=1}^{n_{1}}\,\int\!d^{4}x_{a}\,d^{4}y_{b}\,
\e^{ip_{a}x_{a}-iq_{b}y_{b}} (p_{a}^{2}-m_{1}^{2})(q_{b}^{2}-m_{2}^{2})
\frac{\delta}{\delta j_{1}(x_{a})}\,\frac{\delta}{\delta j_{2}(y_{b})}
W[j] \Big|_{j=0}
\end{array}
\label{2*}
\end{equation}
In the tree approximation one has
\[ W_{tree}[j] = S_{cl}^{(j)}[\fo,\ft] \]
where $\fo[j],\ft[j]$ is the classical solution for the theory with the action

\[ S_{cl}^{(j)} = S_{cl}[\fo, \ft]\,+ \,\int \! d^{4}x\, j_{1} \fo\,+
\int\!d^{4}x\,j_{2}\ft\]
$\fo[j]$ and $\ft[j]$ have to obey the Feynman boundary conditions at
$t\to\pm\infty$.
Notice that
\begin{equation}
\frac{\delta S_{cl}^{(j)}}{\delta j_{i}} = \f_{i}[j]
\label{3*}
\end{equation}

When all particles are at rest, i.e., all ${\bf p}_{a}=0$,
\ ${\bf q}_{b}=0$, the space-time dependent sources
$j_{i}(x)$ and the solutions of the classical field equations
$\f_{i}(x)$ may be replaced by functions $j_{i}(t)$, \
$\f_{i}(t)$ depending only on time, and the classical field
equations become ordinary differential ones. In this limit,
according to eqs.(\ref{2*}) and (\ref{3*}),
the mass-shell amplitudes are obtained by substituting
\[j_{1}(t)=\rho_{1}\,\e^{-i\w_{1}t} ,\,\,\,\,\,\, j_{2}(t)=\rho_{2}
                                           \e^{i\w_{2}t};\]
and then taking the limit $\w_{i}\to m_{i}$ in the
expression
 \begin{equation}
A_{n_{1},\fo \to n_{2},\ft}=(-i)^{n_{1}+n_{2}}\bigl(\w_{1}^{2}-m_{1}^{2}\bigr)^
{n_{1}}\bigl(\w_{2}^{2}-m_{2}^{2}\bigr)^{n_{2}} \frac{\d^{n_{1}-1}}{\d
 \rho_{1}^{n_{1}-1}}\,\frac{\d^{n_{2}}}{\d
\rho_{2}^{n_{2}}}\fo(j,\w)\Big\vert_{j=0}
 \label{3**}
\end{equation}

So, we have to consider the following classical equations,
\begin{eqnarray}
\ddot{\f_{1}}+m_{1}^{2}\f_{1}+2\sum_{k}\la_{1k}\f_{1}\f_{k}^{2}\,
=\,\rho_{1}\e^{-i\w_{1}t}\nonumber\\
\ddot{\f_{2}}+m_{2}^{2}\f_{2}+2\sum_{k}\la_{2k}\f_{2}\f_{k}^{2}\,
=\,\rho_{2}\e^{i\w_{2}t}
\label{6}
\end{eqnarray}
Let us apply the ordinary perturbation technique to this nonlinear
system. To the zeroth order in $\la$ (free theory) the solution is
\begin{eqnarray}
\fo^{(0)}=z_{1}=\dso\e^{-i\w_{1}t};\,\,\,\,\,
\ft^{(0)}=z_{2}=\dst\e^{i\w_{2}t},
\label{4+}
\end{eqnarray}
where
\[\ds_{i}=\frac{\rho_{i}}{m_{i}^{2}-\w_{i}^{2}}\]
At each subsequent step of the iteration procedure, we have to solve the
following
equations,
\begin{equation}
           \ddot{\f}_{i}^{(k)}+m_{i}^{2}\f_{i}^{(k)}
=-\sum_{j}2\la_{ij}(\f_{i}\f_{j}^{2})^{(k-1)}
\label{*}
\end{equation}
where $(\fn\fj^{2})^{(k-1)}$ is of order $\la^{k-1}$ and
therefore is expressed trough
$\fn^{(0)},\fn^{(1)},...\fn^{(k-1)}$. At $\w_{i}\neq m_{i}$,
the perturbative solution is an expansion in $z_{1}$ and
$z_{2}$, whose coefficients are finite in the limit
$\w_{i}\to m_{i}$ until a certain step. At this step (say, $l$-th),
the resonance term (i.e. the term oscillating
with the frequency $\pm m_{i}$ in the limit $\w_{i}\to
m_{i}$) appears for the first time on the right hand side
of eq.(\ref{*}), so that
\[
\ddot{\fo}^{(l)}+m_{1}^{2}\fo^{(l)}\,=\,i\,
\frac{A_{n_{1}n_{2}}(\w_{1},\w_{2})}{(n_{1}-1)!\,n_{2}!}\,
\dso^{n_{1}-1}\dst^{n_{2}}\,\e^{it((n_{1}-1)\w_{1}-n_{2}\w_{2})}+
\mbox{non-resonance terms}
 \]
Up to $\la^{l}$, the solution is
 \begin{equation}
\f(z_{1},z_{2};\w_{1},\w_{2})
=\sum_{(k,r)<(n_{1},n_{2})}C_{kr}(\w)z_{1}^{k}z_{2}^{r}+
\frac{i}{\w_{1}^{2}-m_{1}^{2}}\,
\frac{A_{n_{1}n_{2}}(\w_{1},\w_{2})}{(n_{1}-1)!\,n_{2}!}\,
z_{1}^{n_{1}-1}z_{2}^{n_{2}}
\label{4*}
 \end{equation}
where $C_{kl}(\w)$ and $A_{n_{1}n_{2}}(\w)$ are finite at $\w_{i}\to m_{i}$.
Making use of the definition  of $z_{i}$, eq.(\ref{4+}), we
obtain from eq.(\ref{3**}) the following expression for the
on-shell amplitudes at rest,
\[A=\lim_{\w_{i}\to
m_{i}}(-i)(\w_{1}^{2}-m_{1}^{2})\,\frac{\d^{n_{1}-1}} {\d
z_{1}^{n_{1}-1}}\,\frac{\d^{n_{2}}}{\d z_{2}^{n_{2}}}
\fo(z_{1},z_{2};\,\w_{1},\w_{2})\Big\vert_{z_{i}=0}\]
{}From eq.(\ref{4*}) we see that
\begin{equation}
A=A_{n_{1},n_{2}}(\w_{1},\w_{2})\vert_{\w_{i}=m_{i}}
\label{'}
\end{equation}
So the amplitudes are determined by the resonance term.

For the actual evalution of the on-shell amplitudes at
rest, it is more convenient to study the
{\it sourceless} field equations, instead of the
 system (\ref{6}). To reformulate the procedure, let us
 compare eq.(\ref{4*}) with the perturbative solution of
the sourceless classical equations supplemented by the
following conditions \begin{equation}
\f_{i}^{(0)}=\hat{z}_{i} \label{++} \end{equation} where
\begin{equation} \hat{z}_{1}=\dso\e^{-im_{1}t},\,\,\,\,\,\,
\hat{z}_{2}=\dst\e^{i m_{2}t}
\label{+++}
\end{equation}
The iteration procedure is again determined by eq.(\ref{*}). At the $l$-th step
the resonance term appears
for the first time,
\begin{equation}
\ddot{\fo}^{(l)}+m_{1}^{2}\fo^{(l)}\,=\,i\,
\frac{A_{n_{1}n_{2}}(m_{1},m_{2})}{(n_{1}-1)!\,n_{2}!}\,
\dso^{n_{1}-1}\dst^{n_{2}}\,\e^{im_{1}t}+\mbox{non-resonance terms}
\label{@}
\end{equation}
where $A_{n_{1},n_{2}}(m_{1},m_{2})$ is precisely the coefficient
$A_{n_{1},n_{2}}(\w_{1},\w_{2})$
in eq.(\ref{4*}) taken at $\w_{i}=m_{i}$, i.e. $A_{n_{1},n_{2}}(m_{1},m_{2})$
is the amplitude of the process of interest (see eq.(\ref{'})). The resonance
term in eq.(\ref{@}) gives rise to the peculiar (growing linearly in time)
term in the perturbative solution,
\begin{equation}
\fo^{(l)}=t\,\e^{im_{1}t}
\frac{A_{n_{1}n_{2}}(m_{1},m_{2})}{(n_{1}-1)!\,n_{2}!}\,
\frac{1}{2m_{1}}\dso^{n_{1}-1}
\dst^{n_{2}}+\mbox{oscillating terms}
\label{5*}
\end{equation}

Thus, we obtain the following prescription for
calculating the amplitude $A$ for particles at rest at the tree level.
One has to solve the homogeneous ordinary differential equation --- the
classical equation for space-independent fields --- by the perturbation
technique;
the coeffitient of the first peculiar term multiplied by
$2m_{1}(n_{1}-1)!n_{2}!$ is just equal to the required amplitude.

{\bf 3}. Let us show that there is no peculiar terms in the
perturbative solution to classical equations when the corresponding
mechanical system possesses a special kind of symmetry.

As  an example, let us consider the model with softly broken $O(2)$ symmetry,
determined by eq.(\ref{1*}).
The corresponding classical system, eq.(\ref{1a*}), is the simplest case
of the Garnier system and so is integrable
\cite{Perelomov,Chudn,Grosse,Wojc,Ind'}.
Besides the usual time translation, this system possesses a non-trivial
symmetry
whose infinitesimal form is \cite{Ind}
\begin{eqnarray}
\fo&\mapsto&\tilde{\fo}=\fo+\eps\,\la\ft(\dot{\fo}\ft-\dot{\ft}\fo)\nonumber\\
\ft&\mapsto&\tilde{\ft}=\ft+\eps\bigl[\,\la\fo(\dot{\ft}\fo-\dot{\fo}\ft)
+\frac{m_{1}^{2}-m_{2}^{2}}{2}\dot{\ft}\bigr]
\label{7*}
\end{eqnarray}
where $\eps$ is a small parameter.

To see how this transformation changes the boundary conditions (\ref{++}) and
(\ref{+++}),
we take the limit $\la\to 0$ on the right hand side of eq.(\ref{7*}).
We obtain for the unbroken symmetry case ($<\fo>=<\ft>=0$)
\begin{equation}
\fo^{(0)}\mapsto\tilde{\fo}^{(0)}=\fo^{(0)},\,\,\,\,\,
\ft^{(0)}\mapsto\tilde{\ft}^{(0)}=\ft^{(0)}+
\eps\frac{m_{1}^{2}-m_{2}^{2}}{2}\dot{\ft}^{(0)},
\label{7**}
\end{equation}
so that
\begin{equation}
\tilde{\dso}=\dso,\,\,\,\,\,
\tilde{\dst}=\dst\bigl(1+im_{2}\frac{m_{1}^{2}-m_{2}^{2}}{2}\eps\bigr)
\label{7***}
\end{equation}
The solution $\f_{1,2}$ is determined by the parameters $\dso$, $\dst$;
due to the uniqueness of the perturbative solution, we have the following
identity
\begin{equation}
 \tilde{\f}_{1,2}(\dso,\,\dst)=\f_{1,2}(\tilde{\dso},\,\tilde{\dst})
\label{a}
\end{equation}
where the left hand side is given by eq.(\ref{7*}) and $\tilde{\ds_{i}}$ on the
right hand side are given by eq.(\ref{7***}). Let us compare the peculiar terms
in
this identity. We obtain from eq.(\ref{7*}) that up to order $\la^{l}$
\begin{equation}
\tilde{\fo}(\ds)
 =\fo(\ds)+\mbox{oscillating terms}+O(\la^{l+1})
\label{b}
\end{equation}
while eqs.(\ref{5*}) and (\ref{7***}) give, again up to order $\la^{l}$,
\begin{equation}
\fo(\tilde{\ds})=\fo(\ds)+i\eps t n_{2}\frac{m_{1}^{2}-m_{2}^{2}}{4}
\hat{z}_{1}^{n_{1}-1}\hat{z}_{2}^{n_{2}}\frac{A_{n_{1}n_{2}}}{(n_{1}-1)!
n_{2}!}+\mbox{oscillating terms}+O(\la^{l+1})
\label{c}
\end{equation}
The identity (\ref{a}) is satisfied only when
\[
 (m_{1}^{2}-m_{2}^{2}) A_{n_{1}n_{2}}=0,
\]
i.e. either $A_{n_{1}n_{2}}=0$ or $m_{1}^{2}=m_{2}^{2}$. This means that the
amplitude of the production of $n_{2}$ particles $\ft$ by $n_{1}$ particles
$\fo$, all at rest, does not vanish only for
$n_{1}=n_{2}=2$ (recall that we study the case of coprime $n_{1}/2$ and
$n_{2}/2$).

This analysis can be generalized directly to the broken symmetry case, $m_{1}
^{2}<0$, $m_{2}^{2}>m_{1}^{2}$, when
\begin{equation}
\langle \fo \rangle=\frac{|m_{1}|}{2\sqrt{\la}}
\label{8*}
\end{equation}
and instead of eq.(\ref{7**}) one has
\begin{eqnarray*}
\fo^{(0)}&\mapsto&\tilde{\fo}^{(0)}=\fo^{(0)},\\
\ft^{(0)}&\mapsto&\tilde{\ft}^{(0)}=\ft^{(0)}+\eps\bigl(\frac{|m_{1}|^{2}}{4}-
\frac{|m_{1}|^{2}+m_{2}^{2}}{2}\bigr)\dot{\ft}^{(0)}\\
&& = \ft^{(0)}-\eps\frac{|m_{1}|^{2}+2m_{2}^{2}}{4}\dot{\ft}^{(0)}
\end{eqnarray*}
So,
\begin{eqnarray*}
\tilde{\dso}&=&\dso,\\
\tilde{\dst}&=&\dst\bigl(1-im_{2}\eps\frac{|m_{1}^{2}|+2m_{2}^{2}}{4}\bigr)
\end{eqnarray*}
Eq.(\ref{b}) remains valid for the broken symmetry case, while instead of
eq.(\ref{c})
one has
\[
\fo(\tilde{\ds})=\fo(\ds)+\eps t n_{2}\frac{|m_{1}|^{2}+2m_{2}^{2}}{8}
\hat{z}_{1}^{n_{1}-1}\hat{z}_{2}^{n_{2}}\frac{A_{n_{1}n_{2}}}{(n_{1}-1)!
n_{2}!}+\mbox{oscillating terms}+O(\la^{l+1})
\]
Thus, to satisfy the identity (\ref{a}) one requires
\[(|m_{1}|^{2}+2m_{2}^{2})A_{n_{1}n_{2}}=0\]
Therefore, the amplitudes may not vanish only when $|m_{1}|^{2}+2m_{2}^{2}=0$.
Since
the masses of the excitations around the vacuum (\ref{8*})
are $m_{\fo}=\sqrt{2}|m_{1}|$, $m_{\ft}=\sqrt{|m_{1}|^{2}+m_{2}^{2}}$,
this condition means that $m_{\fo}^{2}=2m_{\ft}^{2}$, i.e. the only
non-vanishing
tree amplitude at rest is that of the decay of a $\fo$-particle into two
$\ft$-particles.

The absence of the peculiar terms in the perturbation series in this model
may be demonstrated by constructing the explicit solution to all orders in
$\la$. For example, in the unbroken case, the solution obeyng the conditions
(\ref{++}) and (\ref{+++}) is (cf. \cite{we})
\[
\fo=\hat{z}_{1}(1-\la\frac{\ka}{2\mt^2}\hat{\zt}^{2})
\Bigl(1-\frac{\la}{2\mo^{2}}\hat{\zo}^{2}-
\frac{\la}{2\mt^{2}}\hat{\zt}^{2}+
\la\frac{\ka^{2}}{4\mo^{2}\mt^{2}}\hat{\zo}^{2}\hat{\zt}^{2}
\Bigr)^{-1}
\]
\[
\ft=\hat{z}_{1}(1+\la\frac{\ka}{2\mo^2}\hat{\zo}^{2})
\Bigl(1-\frac{\la}{2\mo^{2}}\hat{\zo}^{2}-
\frac{\la}{2\mt^{2}}\hat{\zt}^{2}+
\la\frac{\ka^{2}}{4\mo^{2}\mt^{2}}\hat{\zo}^{2}\hat{\zt}^{2}
\Bigr)^{-1}
\]
where
\[
\ka=\frac{\mo+\mt}{\mo-\mt}
\]
Clearly, the expansion of this solution in $\la$ does not contain peculiar
terms.

{\bf 4}. So, we find that the nullification of the tree amplitudes for
$\fo$-particles to create
$\ft$-particles, all particles being at rest and on mass
shell, in the model (\ref{1*}) is directly related
to the non-trivial symmetry of the corresponding Hamiltonian system of
classical mechanics. The only relevant property of this symmetry is that the
expression
for the infinitesimal transformation for at least one of the fields $\f_{i}$
contains a term that is linear in this field or in its derivative.
Thus, in any theory of interacting scalar fields
possessing the symmetry of the kind described above, the on-shell tree
amplitudes
at rest must vanish. Another example of such a model is the
integrable version of H\'enon-Heiles system with arbitrary masses
\cite{Perelomov,Ind}, which corresponds to the field theory with the lagrangian
\begin{equation}
L=\frac{1}{2}(\d_{\mu}{\f}_{1}^{2})+\frac{1}{2}(\d_{\mu}{\f}_{2}^{2})-
\frac{m_{1}^{2}}{2}
\f_{1}^{2}-\frac{m_{2}^{2}}{2}\ft^{2}-\la\fo^{2}\ft-2\la\ft^{3}
\label{d}
\end{equation}
The non-trivial symmetry of the corresponding space-independent
hamiltonian is \cite{Ind}
\begin{eqnarray*}
\fo&\mapsto&\fo+2\la(\fo\dot{\ft}-2\ft\dot{\fo})+
\frac{4m_{1}^{2}-m_{2}^{2}}{2}\dot{\fo},\\
\ft&\mapsto&\ft+2\la\fo\dot{\ft}
\end{eqnarray*}
By repeating the above arguments one finds that the only non-vanishing tree
amplitude at rest in the model (\ref{d}) is that of the decay of one
$\fo$-particle
into two $\ft$-particles.

More examples of the nullification discussed in this paper may be constructed
on
the basis of wider classes of integrable classical systems, some of which can
be found in ref.\cite{Perelomov}.

Of course, to a given order of the perturbation theory, the nullification of
the tree amplitudes at rest for models like (\ref{1*}) or (\ref{d}) can be
seen by explicit evaluation of the Feynman diagrams. In that language,
zeros of the amplitudes emerge as the result of cancellations between various
diagrams
weighted by their symmetry factors. The fact that these zeros are related
to the symmetries of classical integrable systems gives the rationale for
 these, otherwise miraculous, cancellations.

The authors are indebted to D.T. Son, V.P. Spiridonov and P.G. Tinyakov for
 helpful discussions.
 The work of M.L. and S.T.
is supported in part by the Weingart Foundation through a cooperative
agreement with the Department of Physics at UCLA.

\end{document}